\documentstyle[amssymb,preprint,aps]{revtex}

\begin{document}
\title{Universal ultrabright source of entangled photon states:\\
generation and tomographic analysis of Werner states and of Maximally
Entangled Mixed States }
\author{G. Di Nepi, F. De Martini, M. Barbieri and P. Mataloni}
\address{Dipartimento di Fisica and Istituto Nazionale per la Fisica della Materia\\
Universit\`{a} di Roma ''La Sapienza'', Roma, 00185 - Italy}
\maketitle

\begin{abstract}
A novel ultrabright parametric source of polarization entangled photon pairs
with striking spatial characteristics is reported. The distribution of the
output electromagnetic ${\bf k}$-modes excited by Spontaneous Parametris
Down Conversion and coupled to the output detectors can be very broad. It
could coincide with the {\it full set} of phase-matched excited modes, at
least {\it in principle}. In this case a relevant {\it conditional
quasi-pure }output state should be realized. By these (approximate) states
realized over a full {\it Entanglement-Ring} output distribution, the non
local properties of the generated entanglement has been tested by standard
Bell measurements and by Ou-Mandel interferometry. A novel ''{\it %
mode-patchwork}'' technique based on the quantum superposition principle is
adopted to synthesize in a straightforward and reliable way any kind of {\it %
mixed-states, }of large conceptual and technological interest{\it \ }in{\it %
\ }modern Quantum Information. Tunable Werner states and Maximally Entangled
Mixed States (MEMS) have indeed been created by the new technique and
investigated by quantum tomography. A thorough study of the entropic and
nonlocal properties of these states has been undertaken experimentally and
theoretically, by a unifying variational approach. PACS: 03.67.Mn, 03.65.Ta,
03.65.Ud, 03.65.Wj.
\end{abstract}

\section{Introduction}

Entanglement, ''{\it the} {\it characteristic trait of quantum mechanics'' }%
according to Erwin Schroedinger, is playing an increasing role in nowadays
physics \cite{1}. It is the irrevocable signature of quantum nonlocality,
i.e. the scientific paradigm today recognized as the fundamental cornerstone
of our yet uncertain understanding of the Universe. The striking key
process, the ''{\it bolt from the blue}'' [i.e. from the skies of
Copenhagen] according to Leo Rosenfeld \cite{2}, was of course the EPR
discovery in 1935 followed by a much debated endeavour ended, in the last
few decades by the lucky emergence of\ the Bell's inequalities and by their
crucial experimental verification\ \cite{3}. In the last years the violation
of these inequalities has been tested so many times by optical experiments
that (almost) no one today challenges the completeness of quantum
nonlocality.

In the modern science of quantum information (QI)\ entanglement represents
the basis of the exponential parallelism of future quantum computers \cite{4}%
, of quantum teleportation \cite{5,6} and of some kinds of cryptographic
communications \cite{7}.

It appears largely significant that the most successful and reliable
applications of entanglement have been obtained so far in the field of
quantum optics. There the electromagnetic modes are associated with
''qubits'' which are generally encoded by the field polarization $(\pi )\ $%
\cite{8}. This type of qubit encoding is precisely the one considered in the
present work. The $\pi -$entanglement arises within the spontaneous emission
process (Spontaneous Parametric Down Conversion:\ SPDC) in a nonlinear
optical crystal under suitable conditions, as we shall see shortly. In this
process, a pair of correlated photons are generated at wavelengths (wl) $%
\lambda _{1}$ and $\lambda _{2}$ and momenta $\hslash {\bf k}_{1}$ and $%
\hslash {\bf k}_{2}$ by a quantum electrodynamical\ (QED) annihilation of a
pump photon with wl $\lambda _{p}$ and momentum ${\bf k}_{p}$. Conservation
of energy , $\lambda _{1}^{-1}+\lambda _{2}^{-1}=\lambda _{p}^{-1}$ and of
the momentum, i.e. {\it phase-matching }condition ${\bf k}_{1}+{\bf k}_{2}=%
{\bf k}_{p}$ leads to frequency and $k-$vector correlation of the emitted
photons. In condition of entanglement the bipartite Hilbert space $%
H_{1}\otimes H_{2}$ with $dim(H_{1})=dim(H_{2})=2$ is spanned by the four
Bell-state entangled basis, 
\begin{equation}
|\Psi _{\pm }\rangle =2^{-%
{\frac12}%
}\left( |H_{1},V_{2}\rangle \pm |V_{1},H_{2}\rangle \right) \text{, }|\Phi
_{\pm }\rangle =2^{-%
{\frac12}%
}\left( |H_{1},H_{2}\rangle \pm |V_{1},V_{2}\rangle \right)
\end{equation}
where $H$ and $V$ correspond to the horizontal and vertical linear field's
polarizations and the shorthand: $|X_{1},Y_{2}\rangle \equiv |X\rangle
_{1}\otimes |Y\rangle _{2}$ and will be used henceforth. The source of
entanglement adopted today almost exclusively in quantum optics is based on
SPDC in a Type II noncollinear ''phase matched'' nonlinear (NL) crystal in
which a pair of mutually orthogonally polarized photons is generated over a
set of correlated directions corresponding to two different emission cones 
\cite{9}. The cones intersect each other in two particular correlated modes
with k-vectors ${\bf k}_{i}$, $i=1,2$. The overall quantum state emitted
over these two modes may be expressed by either Bell state $|\Psi _{\pm
}\rangle $, depending on a preset NL\ crystal orientation. By a careful
spatial selection of the two correlated k-vectors ${\bf k}_{i}$, e.g. by two
narrow pinholes, a high purity entangled state may be generated by this kind
of source. The typical achievable coincidence count rate is of the order of
few hundreds $\sec ^{-1}$ in a typical case involving a $1mm$ thick NL
crystal excited by a $100mW$ UV pump laser. More recently, Paul Kwiat has
realized a different source of \ $\pi -$entangled pairs, an order of
magnitude brighter than the previous one, based on SPDC emission by a set of
two thin, orthogonally oriented Type I crystal slabs placed stuck in mutual
contact \cite{10}.

In the present work we report on a novel SPDC\ source of
polarization-entangled pairs, recently developed in our laboratory \cite{11}%
, that we believe represents the ultimate solution in the framework of
quantum optics in terms of \ universality and flexibility of generation of
entangled states. The new solution, besides realizing the maximum attainable
''quantum efficiency''\ $(QE)$, i.e. relative ''brightness'', allows the
detection of {\it all} SPDC\ entangled pairs emitted by SPDC over the {\it %
entire} set of wavevectors excited by any parametric scattering process. In
particular, we\ have investigated in the present work the most efficient, $%
\lambda -$degenerate process: $\lambda _{1}=\lambda _{2}=2\lambda _{p}$.
Note briefly in this connection that the present source, when applied to any
Bell inequality test, allows to overcome substantially the\ {\it %
quantum-efficiency}\ ``loophole'', which refers to the lack of {\it detection%
} of \ {\it all couples }of emitted\ correlated photons \cite{12}. This
loophole should indeed be considered to be properly ascribable to the
limited efficiency of the detectors as well as to the loss of the pairs
created in any single QED\ spontaneous emission process that do not reach
the detectors for geometrical reasons. We shall account about this subtle
process in a forthcoming paper dealing with nonlocality tests undertaken by
means of inequalities (Bell-Bohm)\cite{3,13} with and without making
recourse to inequalities (Hardy) \cite{14}.

The present work is mostly devoted to investigate theoretically and
experimentally a quite new branch of QI, the one concerned with the
properties of several relevant families of {\it mixed-entangled states. }%
Because of the unavoidable effects of the decohering interactions indeed
these states are today considered the basic constituents\ of modern QI and
Quantum Computation as they limit the performance of\ all quantum
communication protocols including quantum dense coding{\it \ }\cite{15} and
quantum teleportation \cite{5,6}.{\it \ }It is not surprising that in the
last few years, within an endeavour aimed at the use of {\it mixed-states}
as a practical resource{\it ,} an entire new branch of arduous mathematics
and topology has been created to investigate the unexplored theory of the 
{\it positive-maps} ({\it P-maps}) in Hilbert spaces in view of the
assessment of the ``{\it residual} {\it entanglement''} and of the
establishment of more general ``{\it state}-{\it separability}'' criteria 
\cite{16,17,18}. Very recently this ambitious study has reached results that
are conceptually relevant, as for instance the discovery of a {\it %
discontinuity} in the structure of the{\it \ mixed-state }entanglement{\it . 
}Precisely{\it ,} the identification of\ two classes: the {\it free}-{\it %
entanglement}, useful for quantum communication, and the {\it %
bound-entanglement}, a {\it non-distillable} mysterious process, elicited a
fascinating new horizon implied by the basic question: what is the role of 
{\it bound} entanglement in Nature \cite{19}?

A consistent and wide range investigation of these useful and outmost
appealing aspects of modern physics reqires the availability of a universal,
flexible source by which entangled {\it pure} as well {\it mixed-states} of 
{\it any} structure could be easily enginerered in a reliable and
reproducible way. The novel high-brilliance source described in the present
work indeed possesses these properties and, consistently with the above
considerations, it will be applied first to the realization and to a quantum
tomographic analysis of\ the Werner{\it \ }states with variable mixing
parameters \cite{20,21,22}. At last, the relevant ``{\it maximally entangled
mixed states}'' (MEMS), today of large conceptual and technological
interest, have been easily created and tested by the same techniques \cite
{23}.

The work is organized as follows: in Section II the high-brightness source
is fully described. Moreover the high quality of the realized output
entangled state is demonstrated by a Bell-inequality test showing the
attainment of a nonlocal interferometric visibility as large as $V\geq 0.94$
and a violation of a Bell inequality by 213 standard deviations. Section III
reports on a novel ''patchwork'' method to generate a full set of Werner
states. In addition, the results of a quantum tomographic investigation of
their properties shall be reported. In Section IV the methods of the
previous Section will be extended to a theoretical and experimental study of
the MEMS. In SectionV the EPR\ nonlocality of the MEMS and of the Werner
states shall be taken up in a unifying manner by a variational analysis of
the correlation functions. Finally, the foreseable perspectives of the new
method will be outlined in Section VI.

\section{The high brightness source of entanglement}

The active element of the new source of polarization-entangled photon pairs
consisted of a single Type-I NL thin crystal slab, $\beta $-barium-borate
(BBO),\ excited in two opposite directions $({\bf k}_{p},-{\bf k}_{p})$ by a
UV pump beam with wl $\lambda _{p}$, back-reflected by a spherical mirror $%
(M)$:\ Figure 1. Each of the two equal but independent SPDC processes
generated pairs of correlated photons with wavelengths (wl) $\lambda _{i}$, $%
(i=1,2)$, {\it equal} polarizations $(\pi )$ and with a spatial distribution
of the corresponding, mutually correlated k-vectors ${\bf k}_{i}$ consisting
of two equal, opposite {\it circular} cones (dubbed as ''${\bf k}${\bf -}%
cones''{\bf )}${\bf \ }$with axes ${\bf k}_{p},-{\bf k}_{p}$ \cite{9,10}.
While all the relevant features of the present source are shared by all {\it %
non-degenerate} phase-matched wl configurations, i.e.$\lambda
_{1}^{-1}+\lambda _{2}^{-1}$=$\lambda _{p}^{-1}$ with $\lambda _{1}\neq
\lambda _{2}$, our present demonstration has been (and the present
discussion will be) carried out mostly for the {\it degenerate} case for
simplicity and without loss of generality: $\lambda _{1}=\lambda
_{2}=\lambda =\lambda _{p}/2=727.6nm$. Suppose that the NL\ crystal
orientation is preset as to generate SPDC\ pairs on both ${\bf k}${\bf -}%
cones with any linear polarization, say $\pi =H$. Each pair of each ${\bf k}$%
{\bf -}cone is then emitted with a product state $\left|
H_{1},H_{2}\right\rangle $ over its own $4-$dimensional Hilbert space.
Suppose now that the two ${\bf k}${\bf -}cones are made to overlap into a 
{\it single} one with axis ${\bf k}_{p}$, i.e. directed towards the right
hand side (r.h.s.)\ of Fig.1, by back-reflection over $M$ of the ${\bf k}$%
{\bf -}cone with axis $-{\bf k}_{p}$. If in the flight from the crystal to $%
M $ and back the polarization of the corresponding emitted pairs is flipped
by any unitary transformation, i.e. $H_{i}\rightarrow V_{i}$ and a phase $%
\phi \ $is added, the state of the pair detected on the (r.h.s.)\ of Fig.1
is the {\it entangled} state: 
\begin{equation}
|\Phi \rangle =2^{-%
{\frac12}%
}\left( |H_{1},H_{2}\rangle +e^{i\phi }|V_{1},V_{2}\rangle \right)
\label{phi}
\end{equation}
\ Since this argument holds for {\it any} pair of correlated ${\bf k}_{i}$%
-vectors emitted with equal probability and equal wl $\lambda $, all
diametrally opposite points of the spatial circle obtained by interception
with a plane of the output cone, i.e. belonging to the ''{\it %
entanglement-ring}'' as we shall see shortly, are correlated by the same
entanglement condition and then represented by $|\Phi \rangle $. Of course,
as for all quantum interference phenomena, the ${\bf k}${\bf -}cone
overlapping should be perfect as to realize the ''{\it in principle
indistinguishability}'' of \ the actual origin of the detected pair for any
set of {\it ideal} detectors coupled to the output of the source. All this
represents the key process underlying the present work.

Let us now give more details of the apparatus (Fig. 1). A Type I, $.5mm$
thick, crystal was excited by a slightly focused $V$-polarized cw $Ar^{+}$
laser beam ($\lambda _{p}=363.8nm$) with wavevector $-{\bf k}_{p}$, i.e.
directed towards the l.h.s. in Fig.1. The two photons had common $H$
polarization and were emitted with {\it equal probability\ }over a
corresponding pair of wavevectors belonging to the surface of a cone with
axis $-{\bf k}_{p}$ and aperture $\alpha \simeq 2.9%
{{}^\circ}%
$. The emitted radiation and the laser beam were then back-reflected by a
spherical mirror $M$ \ with curvature radius $R=15cm$, highly reflecting$\ $%
both $\lambda $ and $\lambda _{p}$, placed at a distance $d=R$ from the
crystal. A zero-order $\lambda /4$ waveplate (wp) placed between $M\ $and
the NL\ crystal, i.e. indicated as ''$d-$zone'' in Fig.1, intercepted twice
both back-reflected $\lambda $ and $\lambda _{p}$ beams and then rotated by $%
\pi /2$ the polarization of the back-reflected field with wl $\lambda \ $%
while leaving virtually undisturbed the polarization state of the UV\ pump
beam. In facts, it was verified that the $\lambda /4$\ wp acted closely as a 
$\lambda _{p}/2$ wp since $\lambda _{p}=2\lambda $. The phase $\phi $ $%
(0\leq \phi \leq \pi )$\ of the generated pure entangled state, Eq.(\ref{phi}%
) was reliably controlled by micrometric displacements $\Delta d\ $of $M$\
along ${\bf k}_{p}$. The phase stability, representing the most challenging
experimental problem, was solved by adoption of various tricks, among which
vital was the use of the {\it same} back-mirror $M$ for both wl's $\lambda $
,$\lambda _{p}$ (see Appendix A). A positive lens ($f=15cm$) transformed the
overall emission\ {\it conical }distribution\ into a {\it cylindrical }one
with axis ${\bf k}_{p}$, whose transverse circular section identified the ``%
{\it Entanglement-ring''} ({\it E-ring}) with diameter $D=2\alpha F$. Each
couple of \ points symmetrically opposed through the center of the ring are
then correlated by quantum entanglement, as said. An annular mask with
diameter $D=1.5cm$ and width $\delta =.07cm$ provided in the present
experiment a very accurate spatial selection of the {\it E-ring}. This one
was divided in two equal portions along a vertical axis by a prism-like
two-mirror system and then focused by two lenses on the active surfaces of
two independent photodiodes $A$ and $B$ ({\it Alice} and {\it Bob}) by which
standard Bell measurements could be performed by means of polarization
analyzers ($\overrightarrow{\pi }$). The detectors were silicon-avalanche
mod. SPCM-AQR14 with quantum efficiency $QE=65\%$ and dark count rate $%
\simeq 50s^{-1}$. Typically, two equal interference filters,\ placed in
front of the $A$ and $B$ detectors, with bandwidth $\Delta \lambda =6nm$,
determined the {\it coherence-time }of the emitted photons:\ $\tau _{coh}$%
{\it \ }$\approx 140fsec.$ Optionally, the focusing could have been done on
the ends of two optical fibers that could convey the radiation to two far
apart detection stations $A$ and $B$.

The insertion of one (or several) waveplates of filters partially (or fully)
intercepting the radiation associated with the ${\bf k}${\bf -}cones in (or
ouside) the $d$-zone allowed the realization of \ various configurations of
relevant physical significance. For instance, the insertion of a zero-order $%
\lambda /2$ wp in one of the output detection arms allowed to locally
transform Eq.(\ref{phi}) into the state:

\begin{equation}
|\Psi \rangle =2^{-%
{\frac12}%
}\left( |H_{1},V_{2}\rangle +e^{i\phi }|V_{1},H_{2}\rangle \right)
\label{psi}
\end{equation}
Then, by easily setting $\phi =0\ $\ of $\phi =\pi $ Eq.1 could be locally
transformed in any one of the remaining three Bell states. Furthermore and
most important, the insertion of a zero-order $\lambda _{p}/4$ wp in the $d$%
-zone and intercepting {\it only }the UV\ beam allowed the engeneering of
tunable {\it non maximally entangled} states. As shown by Lucien Hardy, by
this class of states expressed as: 
\begin{equation}
|\Phi \rangle =\alpha |H_{1},H_{2}\rangle +\beta |V_{1},V_{2}\rangle ;\text{ 
}\alpha \neq \beta ;\text{\ }\left| \alpha \right| ^{2}+\left| \beta \right|
^{2}=1
\end{equation}
the nonlocality of quantum mechanics could be tested without recourse to
inequalities (''Hardy's ladder proof'') \cite{14}. In our system, the
rotation of the UV\ wp by an angle $\theta _{p}$\ determines a $\pi $%
-rotation of the back-reflected UV\ pump beam respect to the optical axes of
the (fixed) NL\ crystal slab. The consequent $\theta _{p}$-tunable
unbalancement of the SPDC\ efficiencies over the two ${\bf k}${\bf -}cones
affects the two interfering product-state terms in Eq.(\ref{phi}) by a
coefficient $\propto \cos ^{2}2\theta _{p}$. By adjusting $\theta _{p}$ in
the range $0-\pi /4$, the {\it degree of entanglement }$\gamma =\left|
\alpha /\beta \right| $ can be continuously tuned between $1$ and $0$.
Successful, preliminary results of the Hardy's ladder proof \ obtained by
adoption of the present source have been reported in \cite{11}. More
complete data will be reported elsewhere.

All these consideration are but a preliminary demonstration of the extreme
flexibility of the present source. In the following Sections we shall learn
how, by a simple ''{\it patchwork}'' technique approximately ''{\it pure}''
states can be easily transformed into ''{\it mixed}'' states with various
degree of mixedeness. This allowed us to undertake a comprehensive study of
relevant quantum states as the Werner states and the Maximally Entangled
Mixed States (MEMS).

\subsection{Generation of approximate conditionally pure-states}

The structural characteristics of the quantum state of any photon pair
generated by our source should be analyzed by accounting first for the
excited electromagnetic (e.m.) modes which, in our case are grouped in
correlated pairs by the $3-$wave SPDC interaction. Assume that each SPDC $%
{\bf k}${\bf -}cone is represented by a linear superposition of $N\gg 1$
correlated pairs of e.m. mode sets, ${\bf k}_{ij}\equiv \left\{ {\bf k}_{1j},%
{\bf k}_{2j}\right\} $ where each component is a {\it plane-wave} mode
represented by a $k$-vector ${\bf k}_{j}$. The symbol $j$, ranging from $1$
to $N$ represents the {\it full set} of \ e.m. modes that are coupled to any
elementary QED $3-$wave parametric scattering process taking place in the
NL\ crystal and whose excitation is dynamically allowed by the {\it %
phase-matching} conditions. The symbol $i=1,2$ accounts, as previously, for
the spin-$%
{\frac12}%
$ Hilbert space finally coupled to the detectors $A$ or $B$, respectively.
Since only {\it one pair} of photons is detected at the output of the
source, each of these ${\bf k}_{ij}$ mode sets corresponds to a Fock $2-$%
mode product-state that can be either $|0,0\rangle _{j}$ or $%
|1_{H},1_{H}\rangle _{j}$ or $|1_{V},1_{V}\rangle _{j}$. These two last
product-states correspond to the states expressed in a less precise form in
Eq.(\ref{phi}). Finally, owing to the linearity of quantum mechanics, the
overall {\it entangled-state} expressing a single pair emission and detected
at the output of our source after ${\bf k}${\bf -}cone overlapping and $\phi
-$phase delay may finally be expressed by the quantum superposition: 
\begin{equation}
|\Phi \rangle =2^{-%
{\frac12}%
}%
\mathop{\textstyle\sum}%
{N \atop j:1}%
\ \left( |1_{H},1_{H}\rangle _{j}+e^{i\phi }|1_{V},1_{V}\rangle _{j}\right)
\label{sum}
\end{equation}
\ over a set of plane-wave modes, assumed here {\it discrete} for
simplicity. This should be indeed the {\it exact} form of the output state
of the source, if the{\it \ full set} of $j$-modes could be coupled to the
cathode of the detectors $A$ and $B$. Indeed, it is not difficult to find
that this full coupling scheme is virtually {\it made possible} by the
actual structure of our source in all possible wavelength configurations,
either {\it degenerate} $(\lambda _{1}=\lambda _{2})$ or {\it non-degenerate}
$(\lambda _{1}\neq \lambda _{2})$. In other words, an experiment may be
conceived in principle (an approximate one is in fact in progress in our
laboratory) by which the {\it full set} of QED\ excited modes at any
wavelength, ranging from very large $\lambda _{i}$ down to $\lambda
_{i}=\lambda _{p}$ can be coupled to the detectors {\it without} any
geometrical constraint or spatial or $\lambda $-filtering. Of course, in
this case severe limitations for a {\it full} {\it particle} {\it detection}
come from the limited $\lambda -$extension of the photocathode $QE^{\prime
}s $\ and of the performance of \ the optical components (mirrors, lenses
etc.). Nevertheless, this does not affect {\it in principle} the {\it %
structural character} of the output entangled-state.

Note that in all SPDC-based quantum optics experiment carried out so far,
the set of ${\bf k}_{ij}$ modes coupled to the detectors was drastically
reduced by the use of narrow spatial-filtering pinholes\ \cite{6,8,9,10}.
Actually, the goal there was the realization of particle detection over a 
{\it single} pair of correlated ${\bf k}-$vectors, i.e.only one $j$ term of
the distribution appearing in Eq.(\ref{sum}). As a consequence, in all
experiments performed thus far the drastically truncated sum in Eq.5 implied
necessarily a {\it highly} {\it mixed} character of the output state. These
considerations lead to at least two important consequences:

A) The{\it \ high-brilliance }of the source, due to the full output mode
coupling was found experimentally to $\simeq 2$ order of magnitudes larger
than the conventional one based on SPDC excited by a UV laser with the same
intensity.

B) The {\it quasi-purity} of the generated output state may be considered as
follows. The well known {\it unitary} character of the quantum operator $%
\hat{S}$\ accounting for\ SPDC assures that the purity of the {\it input
state} implies also the purity of the {\it output state: }$|\Phi \rangle
_{out}=\hat{S}|\Phi \rangle _{in}$ \cite{25}. Adopting the common hypotesis
of a undepleted ''classical'' pump beam, the input\ {\it pure state} is
expressed by the overall vacuum-state character of the input modes of the $%
3- $wave parametric interaction. Within the single-pair emission
approximation, the output (pure)\ state $|\Phi \rangle _{out}$ consists of
the\ sum of the state $|\Phi \rangle $, given by Eq.(\ref{sum}), and of the
vacuum-state expressing the {\it non realization} of the QED parametric
scattering process in the NL\ crystal. As a consequence, $|\Phi \rangle $
expressed by Eq.(\ref{sum}) {\it is not, }strictu sensu{\it ,} a pure state
but one out of a two components {\it mixture}.\ However, in the actual case
of a {\it conditional experiment, }i.e{\it . }where the overall detection
system is activated by a trigger pulse elicited by the source itself, the
output vacuum-state contribution can be eliminated. In\ this case the output
state $|\Phi \rangle $ given by Eq. (\ref{phi}) may be considered a {\it pure%
} one. This last condition is usually referred to as expressing the ''{\it %
conditional purity}'' of the output state.

\subsection{Bell inequalities test}

The $|\Phi _{-}\rangle $ Bell state expressed by Eq.2 with $\phi =\pi $,
i.e. a ''singlet'' has been adopted to test the violation of a Bell
inequality by the standard coincidence technique and by the experimental
configuration shown in Figure 1 \cite{26}. According to the previous
considerations, the distribution of active e.m. modes\ appearing in Eq. (\ref
{sum}) and corresponding to the whole {\it E}-{\it ring }was
spatially-filtered{\it \ }by the annular mask, as said. In spite of this\
drastically limiting cutoff we refer to the output state as a {\it quasi-pure%
} state.

The adopted angle orientations of the $\overrightarrow{\pi }$-analyzers
located at the $A$ $(1)$ and $B$ $(2)$ sites were: $\left\{ \theta
_{1}=0,\theta _{1}^{\prime }=45%
{{}^\circ}%
\right\} $ and $\left\{ \theta _{2}=22.5%
{{}^\circ}%
,\theta _{2}^{\prime }=67.5%
{{}^\circ}%
\right\} $, together with the respective orthogonal angles: $\left\{ \theta
_{1}^{\bot },\theta _{1}^{^{\prime }\bot }\right\} $ and $\left\{ \theta
_{2}^{\bot },\theta _{2}^{^{\prime }\bot }\right\} $. By these values, the
standard Bell-inequality parameter could be evaluated \cite{27}: 
\begin{equation}
S=\left| P\left( \theta _{1},\theta _{2}\right) -P(\theta _{1},\theta
_{2}^{\prime })+P\left( \theta _{1}^{\prime },\theta _{2}\right) +P\left(
\theta _{1}^{\prime },\theta _{2}^{\prime }\right) \right|
\end{equation}
where 
\[
P\left( \theta _{1},\theta _{2}\right) =\frac{[C\left( \theta _{1},\theta
_{2}\right) +C\left( \theta _{1}^{\bot },\theta _{2}^{\bot }\right) -C\left(
\theta _{1},\theta _{2}^{\bot }\right) -C\left( \theta _{1}^{\bot },\theta
_{2}\right) ]}{[C\left( \theta _{1},\theta _{2}\right) +C\left( \theta
_{1}^{\bot },\theta _{2}^{\bot }\right) +C\left( \theta _{1},\theta
_{2}^{\bot }\right) +C\left( \theta _{1}^{\bot },\theta _{2}\right) ]} 
\]
and $C\left( \theta _{1},\theta _{2}\right) $ is the coincidence rate
measured at sites $A${\it \ }and ${\it B}$. The measured value $S=2.5564\pm
.0026$ \cite{11}, obtained by integrating the data over $180s$, corresponds
to a violation as large as $213$ standard deviations respect to the limit
value $S=2\ $implied by local realistic{\it \ }theories{\it \ }\cite{3,27,28}%
.As for the full set of measurements reported in the present paper, the good
performance was, of course partially attributable to the high brightness of
the source by which a large set of statistical data could be accumulated in
very short measurement times and with a low UV\ pump power.

The experimental data given in Figure 2 show the$\ \overrightarrow{\pi }$%
-correlation obtained\ by varying the angle $\theta _{1}$ in the range ($45%
{{}^\circ}%
-135%
{{}^\circ}%
$), having kept fixed the angle $\theta _{2}=45%
{{}^\circ}%
$. The interference pattern shows the high degree of $\pi $-entanglement of
the source. The measured visibility of the coincidence rate, $V\geq 94\%$,
gives a further strong indication of the entangled nature of the state over
the entire cone emission ${\bf k}${\bf -}cone, while the single count rates
don't show any periodical fringe behaviour as expected. In the same Fig.2
the dotted line corresponds to the limit boundary between the quantum and
the classical regimes \cite{27} while the theoretical continuous curve
expresses the {\it ideal} interferometric pattern with maximum visibility%
{\it :} $V=1$.

The entanglement character of the source has been investigated by a standard
Ou-Mandel interferometric test \cite{11,hong}, in this case with the Bell
state expressed by Eq.(\ref{psi}): Fig.3, inset. For this experiment the
radius of the iris diaphragms ($ID$) (Fig. 1) was set at $r=0.75mm$. The
experimental results, shown in Fig. 3 and corresponding to the values $\phi
= $ $\pi $ and $\phi =0$, give a value of interference visibility: $V\simeq
88\%$. The FWHM ($\simeq 35\mu m$) of the interference pattern is in good
agreement with the expected value evaluated for a filter bandwidth $\Delta
\lambda =6nm$. In the same Fig. 3 the experimental results corresponding to
the non-interference case $\phi =\pi /2$ are also shown for comparison.

We have characterized the robustness and the brightness of the source by
measuring coincidences for different values of radius $r$\ of the ($ID$).
This corresponds to select different portions of the {\it E-ring}, with
area\ ${\cal S}=2D\delta \arcsin (\frac{r}{D})$ (Fig. 1, inset). The
experimental results of Fig. 4 demonstrate that a coincidence rate larger
than $4\times 10^{3}\sec ^{-1}$ at a pump power $P\simeq 100mW$ is measured
over the entire {\it E-ring} with a still relevant value of visibility. This
outperforms the overall collection efficiency of the SPDC\ process.\ By
taking into account the UV pump power and the overall efficiency of the
apparatus (transmission of the optical plates + detector $\ QE^{\prime }s$),
we can evaluate that the entangled photon pairs are generated at a rate
larger than $2\cdot 10^{5}s^{-1}$. On the other hand, because of the
continuous wave excitation regime, the NL\ parametric gain is so small, $%
g<10^{-3}\ $that the ratio of the probabilities of simultaneous SPDC\
generation of 2 photon pairs and of a single pair is $\leq 10^{-6}$. As a
consequence, the emission of a double pair is negligible with cw pumping.

The present demonstration has been carried out in the $\lambda $-degenerate
condition, i.e. implying the largest QED emission probability according to
NL\ Optics. Note however, as said, that the apparatus works, without any
structural change for a very general $\lambda $-non-degenerate dynamics. \
This would allow, at least in principle and by the use of sufficiently
broadband detectors, the simultaneous detection of \ the (almost) complete 
{\it set} of \ SPDC generated pairs.

Note the high structural flexibility of this novel SPDC\ source. Its
structure indeed suggests the actual implementation of several relevant
schemes of quantum information and communication, including entanglement
multiplexing, joint entanglement over $\overrightarrow{\pi }$ and $k$-vector
degrees of freedom etc. Furthermore, it also suggests the realization of a
spherical cavity Optical Parametric Oscillator (OPO)\ emitting a non-thermal,%
{\it \ E-ring} distribution of entangled photon states. Some od these ideas
are presently being investigated in our Laboratory.

\section{Generation and tomographic characterization of Werner\ states}

Because of the peculiar spatial superposition property of the output state
shown by the structure of Eq. (\ref{sum}), the present apparatus appears to
be an ideal source of {\it any }bi-partite, two-qubit entangled state,
either {\it pure} or {\it mixed}. In particular of\ the Werner state \cite
{20}: 
\begin{equation}
\rho _{W}=p|\Psi _{-}\rangle \left\langle \Psi _{-}\right| +\frac{1-p}{4}%
{\bf I}_{4}
\end{equation}
consisting of a mixture of a {\it pure} singlet state $|\Psi _{-}\rangle
=2^{-%
{\frac12}%
}\left\{ \left| HV\right\rangle -\left| VH\right\rangle \right\} $ with
probability $p$ ($0\leq p\leq 1$) and of a fully {\it mixed-state }expressed
by the unit operator ${\bf I}_{4}$ defined in the $4-$dimensional Hilbert
space. The corresponding density matrix, expressed in the basis $|HH\rangle $%
, $|HV\rangle $, $|VH\rangle $, $|VV\rangle $ is: 
\begin{mathletters}
\begin{equation}
\rho _{W}=\left( 
\begin{array}{cccc}
A & 0 & 0 & 0 \\ 
0 & B & C & 0 \\ 
0 & C & B & 0 \\ 
0 & 0 & 0 & D
\end{array}
\right)  \eqnum{8}
\end{equation}
with:\ $A$=$D$=$%
{\frac14}%
(1-p)$, $B$=$%
{\frac14}%
(1+p)$, $C$=$-p/2$. The Werner states possess a highly conceptual and
historical value because, in the probability range $[1/3<p<1/\sqrt{2}]$,
they {\it do not} violate any Bell's inequality in spite of being in this
range {\it nonseparable} entangled states, precisely {\it NPT states} \cite
{17}.

How to syntezize by our source these paradigmatic, utterly remarkable states
?

Among many possible alternatives, we selected a convenient \ {\it patchwork }
technique based again on the quantum superposition principle. \ This
procedure requires only the three optical elements shown in the grey regions
of Fig. 1. They were arranged in the experimental layout according to the
following steps:

${\bf [1]}$ Making reference to the original {\it source-state} expressed by
Eq.(\ref{phi}), a {\it singlet }state $|\Psi _{-}\rangle $ was easily
obtained by inserting a $\overrightarrow{\pi }${\it -flipping,} zero-order $%
\lambda /2$ wp in front of detector $B.$

$[{\bf 2}]${\bf \ }A anti-reflection coated glass-plate ${\it G}$, $200\mu
m\ $thick, inserted in the $d-section$ with a variable trasverse position $%
\Delta x$, introduced a decohering fixed time-delay $\Delta t>\tau _{coh}\ $%
that spoiled the{\it \ in-priciple\ indistinguishability} of the {\it %
intercepted portions} of the overlapping {\it quantum-interfering} radiation
cones, represented by the ${\bf B}+{\bf C}$ sectors of the {\it E-ring} in
Fig.5. The distinguishability issue is easily solved by thinking that the
induced $\Delta t-$delay allows, {\it in principle} an {\it ideal} detector,
i.e. with infinite time resolution, to identify the ${\bf k}${\bf -}cone
from which the detected particle came from.

As a consequenced in the intercepted sector ${\bf B}+{\bf C,}$ the
statistical mixture $\frac{1}{2}\left[ \left| H_{1},V_{2}\right\rangle
\left\langle H_{1},V_{2}\right| +\left| V_{1},H_{2}\right\rangle
\left\langle V_{1},H_{2}\right| \right] $ was generated,{\it \ }while the
non intercepted sector ${\bf A}$ expresses the polarization {\it pure-state}
singlet contribution to $\rho _{W}$. In other words, {\it all nondiagonal} \
elements of $\rho _{W}$ contributed by the surface sectors ${\bf B}+{\bf C}$
of the E-ring, the ones optically intercepted by ${\it G}$, were set to {\it %
zero }while the non intercepted sector ${\bf A}$ expressed the {\it %
pure-state} singlet contribution to $\rho _{W}$.

${\bf [3]}$ A $\lambda /2$ wp was inserted in the semi-cylindrical photon
distribution reflected by the beam-splitting prism towards the detector $A$.
Its position was carefully adjusted in order to intercept {\it half }of the $%
{\bf B}+{\bf C}\ $sector, i.e. by making ${\bf B}={\bf C}$. Note that only%
{\it \ half }of the E-ring needed to be intercepted by the optical plates,
in virtue of the EPR\ nonlocality. In summary, the sector ${\bf A}$ of the 
{\it E-ring} contributes to $\rho _{W}\ $with a {\it pure} state $p|\Psi
_{-}\rangle \left\langle \Psi _{-}\right| $, the sector ${\bf B}+{\bf C=2B}$
with the statistical mixture: $%
{\displaystyle{1-p \over 4}}%
\left\{ \left[ \left| H_{1},V_{2}\right\rangle \left\langle
H_{1},V_{2}\right| +\left| V_{1},H_{2}\right\rangle \left\langle
V_{1},H_{2}\right| \right] +\left[ \left| H_{1},H_{2}\right\rangle
\left\langle H_{1},H_{2}\right| +\left| V_{1},V_{2}\right\rangle
\left\langle V_{1},V_{2}\right| \right] \right\} $ and the probability $p,$
a monotonic function of $\Delta x$ ($p\propto \Delta x$ for small $p$),\
could be easily varied over its full range of values, from $p=0$ ($\rho _{W}=%
\frac{1}{4}{\bf I}_{4}$) to $p=1$ ($\rho _{W}=\left| \Psi _{-}\right\rangle
\left\langle \Psi _{-}\right| $) \cite{29}. Optionally, the setting of the $%
\lambda /2$ wp intercepting the beam towards $A{\cal \ }$could be
automatically activated by the {\it single} setting $\Delta x$, e.g. via an
electromechanical servo.As an example, we may give a detailed demonstration
of the feasibility of our technique by synthesizing the identity matrix\ $%
{\bf I}$ (Fig. 6). First insert the $\lambda /2$ wp on channel $A$ to
intercept $\frac{1}{4}$ of the {\it E-ring: }in this \ way the mixture $%
\frac{1}{2}\left[ \left| \Phi _{-}\right\rangle \left\langle \Phi
_{-}\right| +\left| \Psi _{-}\right\rangle \left\langle \Psi _{-}\right| %
\right] $\ is produced (Fig 6a). The complete erasure of {\it non diagonal}
elements of the matrix is then performed by making a $\Delta t-$delay glass
plate ${\it G}$ to intercept half of the {\it E-ring} (Fig. 6b)

{\it Any }Werner state could be realized by a similar {\it patchwork}
technique,\ by setting ${\bf B}={\bf C}\ $and by adjusting the value of $\
p(\Delta x)$. Far more generally, {\it all possible }bi-partite states in $%
2\times 2\ $dimensions could be created by this technique, as we shall see
shortly. This indeed expresses the\ ''{\it universality''} of our source.

The tomographic reconstructions \cite{21} of three different Werner states
are shown in Fig. 7, with weight $p=$ $0.82$ (a), $p=0.47$ (b) and $p=$ $%
0.27 $ (c). The imaginary components are negligible.

As anticipated at the beginning of the present Section, the Werner states
have been introduced as examples of non separable states that, in a proper
range of $p$, do not violate CHSH inequality. These states are also
important for quantum information, since they model a decoherence process
occurring on a singlet state travelling along a noisy channel \cite{22}.

We may investigate these processes with some details on the basis of some
relevant entropic properties of the mixed states. A relevant property of any
mixed-state, the ''{\it tangle}''$T=[C(\rho )]^{2}$, i.e. the square of \
the {\it concurrence} $C(\rho )$,$\ $is directly related to the {\it %
entanglement of formation} $E_{F}(\rho )\ $and expresses the degree of
entanglement of $\rho $ \cite{30}. Another important property of the
mixed-states is the ''{\it linear entropy'' }$S_{L}=d(1-Tr\rho ^{2})/(d-1)$, 
$S_{L}=(1-p^{2})$ for the Werner states, which quantifies the degree of
disorder, viz. the {\it mixedeness} of a system with dimensions $d$ \cite
{TSl}. In virtue of the very definition of $C(\rho )$, these two quantities
are found to be related, for the Werner states, as follows: 
\end{mathletters}
\begin{equation}
T_{W}\left( S_{L}\right) (=\left\{ 
\begin{array}{c}
{\frac14}%
(1-3\sqrt{1-S_{L}})^{2}\text{ for }0\leq S_{L}\leq 
{\displaystyle{8 \over 9}}%
\text{,} \\ 
0\text{ \ \ \ \ \ \ \ \ \ \ \ \ \ \ \ \ \ \ \ \ \ \ \ \ for }%
{\displaystyle{8 \over 9}}%
\leq S_{L}\leq 1\text{.}
\end{array}
\right.  \label{wer}
\end{equation}
Since $T=0$ iff the state is separable, we deduce that the Werner states are
not separable in the range $0\leq S_{L}<%
{\displaystyle{8 \over 9}}%
$ or, equivalently, in the range $%
{\displaystyle{1 \over 3}}%
<p\leq 1$. Several Werner states have been generated by this technique and
relative experimental values are plotted in the $T$ vs. $S_{L}$ plane shown
in Fig. 8 \cite{31}. The agreement between experimental data and theoretical
curve appears to be good.

We have recently adopted Werner states generated by this technique to
perform the first experimental realization of the ''{\it entanglement
witness''}, a poweful method to detect entanglement with few local
measurements \cite{32}.

\section{Generation and tomographic characterization of MEMS}

As a final demonstration of the universality of our method, a\ full set of \
Maximally entangled Mixed States (MEMS) has been synthesized by our source
and tested again by quantum tomography \cite{21,23,MEMS}. On the other hand,
according to the introductory notes expressed above, the MEMS are to be
considered, for practical reasons, as peculiar ingredients of modern quantum
information because their entanglement can not be increased by any unitary
transformation. Since the Werner states share this property they can be
assumed to belong to the broader class of MEMS.

This last statement can be proved \cite{34}\ by expressing the density
matrix $\rho _{W}$ in terms of the {\it fidelity} $F$ representing the
overlapping between any Werner state and $\left| \Psi _{-}\right\rangle
\left\langle \Psi _{-}\right| $: 
\begin{equation}
\rho _{W}=\frac{1-F}{3}{\bf I}_{4}+\frac{4F-1}{3}\left| \Psi
_{-}\right\rangle \left\langle \Psi _{-}\right| \text{.}  \label{fidelity}
\end{equation}
Hence the expression for $T$ is given by: 
\begin{equation}
T\left( \rho _{W}\right) =\left( \max \left\{ 0,2F-1\right\} \right) ^{2}
\end{equation}
The condition for an entangled state implies that $%
{\frac12}%
<F\leq 1$, and $T\left( \rho _{W}\right) =\left( 2F-1\right) ^{2}$. A
nonlocal unitary transformation expressed in terms of the parameter $a$,
where $a\in \left[ 
{\frac12}%
,1\right] $%
\begin{equation}
U\left| \Psi _{-}\right\rangle =\left| \Psi \right\rangle =\sqrt{a}\left|
H_{1},H_{2}\right\rangle +\sqrt{1-a}\left| V_{1},V_{2}\right\rangle
\end{equation}
transforms $\rho _{W}$ into the mixture: 
\begin{equation}
\rho _{W}^{\prime }=U\rho _{W}U^{\dagger }=\frac{1-F}{3}I_{4}+\frac{4F-1}{3}%
\left| \Psi \right\rangle \left\langle \Psi \right| \text{.}
\end{equation}
It has been demonstrated that $\rho _{W}^{\prime }$ is still entangled for $%
{\frac12}%
\leq a<%
{\frac12}%
\left( 1+%
{\displaystyle{\sqrt{3\left( 4F^{2}-1\right) } \over 4F-1}}%
\right) $. $T\left( \rho _{W}\right) $ is then maximized for $a=%
{\frac12}%
$. As a consequence, any Werner state belongs to the class of MEMS.

The MEMS generated and tested by our method are the ones that own {\it %
maximum tangle} allowed for a given value of linear entropy. The density
matrix $\rho _{MEMS}$ is represented by the matrix given by Eq. (8) with the
parameters: $A\equiv (1-2g(p))$, $B\equiv g(p)$, $C\equiv -%
{\frac12}%
p$, $D\equiv 0$. There: $g(p)=%
{\frac12}%
p$\ for $p\geq 
{\displaystyle{2 \over 3}}%
$ and $g(p)=%
{\displaystyle{1 \over 3}}%
$ for $p<%
{\displaystyle{2 \over 3}}%
$. The expression of $T$, obtained by a procedure similar to as the one
adopted for Werner states, is 
\begin{equation}
T_{MEMS}(S_{L})=\left\{ 
\begin{array}{c}
{\frac14}%
(1+\sqrt{1-%
{\displaystyle{3 \over 2}}%
S_{L}})^{2}\text{, \ \ for }0\leq S_{L}\leq 
{\displaystyle{16 \over 27}}%
\\ 
{\displaystyle{4 \over 3}}%
-%
{\displaystyle{3 \over 2}}%
S_{L}\text{, \ \ \ \ \ \ \ \ \ \ \ \ \ \ \ \ \ \ \ for}\ 
{\displaystyle{16 \over 27}}%
\leq S_{L}\leq 
{\displaystyle{8 \over 9}}%
\text{.}
\end{array}
\right.  \label{MEMS}
\end{equation}

Two different partition techniques of the {\it E-ring} have been adopted to
generate\ the MEMS, depending on the singlet weight, either $p<%
{\displaystyle{2 \over 3}}%
$ or $p\geq 
{\displaystyle{2 \over 3}}%
$. Let's consider three different experimental steps for $p<%
{\displaystyle{2 \over 3}}%
$.

${\bf [1]}$ The $\lambda /2$ wp was inserted in the semi-cylindrical photon
distribution reflected by the beam-splitting prism towards the detector $A$
in order to divide in equal sectors, the two rings corresponding to the left 
${\bf k}$-cone (${\bf D}={\bf E+F}$, Fig. 12a) and to the right ${\bf k}$%
-cone (${\bf G}={\bf H}$, Fig. 9b). This generated at the output the
statistical mixture: $\left| V_{1},V_{2}\right\rangle \left\langle
V_{1},V_{2}\right| $, $\left| V_{1},H_{2}\right\rangle \left\langle
V_{1},H_{2}\right| $, $\left| H_{1},H_{2}\right\rangle \left\langle
H_{1},H_{2}\right| $, $\left| H_{1},V_{2}\right\rangle \left\langle
H_{1},V_{2}\right| $.

${\bf [2]}$ Sector ${\bf D}$ of the left ${\bf k}$-cone, corresponding to $%
\left| V_{1},V_{2}\right\rangle \left\langle V_{1},V_{2}\right| $, was
erased by inserting a right-angle opaque screen (Fig. 9a), while sector $%
{\bf G}$ of the right ${\bf k}$-cone, corresponding to $\left|
H_{1},H_{2}\right\rangle \left\langle H_{1},H_{2}\right| $ was kept
unaltered (Fig. 9b).

$\left[ {\bf 3}\right] $ The glass-plate ${\it G}$ was inserted in the $%
d-section$ of the source in order to intercept a portion $1-p$\ of the
sector ${\bf E}$ of the left ${\bf k}$-cone, ${\bf F}$ in Fig. 9a. As for
the generation of the Werner states, this procedure spoiled the{\it \ }%
indistinguishability of the intercepted portions of the overlapping
radiation cones. The three identical non-zero diagonal terms, corresponding
to sector ${\bf G}$, and to the sectors overlapped with ${\bf H}$, ${\bf E}$
and ${\bf F}$ were: $\left| H_{1}H_{2}\right\rangle \left\langle
H_{1}H_{2}\right| $, $\left| V_{1}H_{2}\right\rangle \left\langle
V_{1}H_{2}\right| $, $\left| H_{1}V_{2}\right\rangle \left\langle
H_{1}V_{2}\right| $, for $p$ varying in the range $0\leq p<%
{\displaystyle{2 \over 3}}%
$ by transverse displacement of ${\it G}$. In this way the ratio between the 
{\it E-ring} contributions ${\bf E}$ and ${\bf F}$ in the left ${\bf k}$%
-cone could be continuously tuned. The larger the ${\bf F}$ contribution,
the larger was the decoherence of the MEMS.

In the case $p\geq 
{\displaystyle{2 \over 3}}%
$, no retarding glass plate was needed to realize the MEMS. By fine
adjustment of the $\lambda /2$ wp the weight $p$ of the {\it singlet} could
be tuned in the range $%
{\displaystyle{2 \over 3}}%
\leq p<1$ and, for each value of $p$ the vertical position of the screen
needed to be adjusted in order to erase the contribution $\left|
V_{1},V_{2}\right\rangle \left\langle V_{1},V_{2}\right| $ of sector ${\bf D}
$ in Fig. 9c.

An accurate experimental production of the MEMS was found particularly
severe likely because of the critical requirements needed for operating on
the very boundary between the {\it allowed} and the {\it forbidden} region
of the $(S_{L},T)$\ plane \cite{31}\ \ In this sense any lack of correlation
within the singlet contribution limited the quality of MEMS. In our source
the BBO crystal axis was oriented in a vertical plane and a strong
decoherence effect could arise from spatial walkoff occurring in the NL\
crystal between the vertically polarized twin photons belonging to the left $%
{\bf k}$-cone. This one was associated with the extraordinary index of
refraction $n_{e}(\zeta )$ of the NL crystal, where $\zeta $\ is.the angle
between the ${\bf k}$ wv of each photon and the axis direction. This effect,
which is also present in the $\lambda /4$ wp, may be reduced by working with
a small aperture angle $\alpha $ of the emission cone. By varying the BBO
axis orientation, in the experiment the MEMS were produced by reducing the
cone aperture to the value $\alpha \simeq 1.4%
{{}^\circ}%
$ .

The tomographic results shown in Fig.10 reproduces graphically, and with
fair accuracy two $\rho _{MEMS\;}$structures with parameters\ $p=0.77$\ (a)
and $p=0.45$ (b), while Fig. 11 shows the experimental behaviour of several
experimental MEMS in the $(S_{L},T)$\ plane compared with the theoretical
curves of Eqq. (\ref{wer}) and (\ref{MEMS}). The produced states lie closely
to the theoretical curve for MEMS, however the agreement between the
experimental results and the theoretical predictions was found at last less
satisfactory than for Werner states.

\section{Non local properties of MEMS}

It is possible to investigate the non local properties of MEMS assuming a
general formalism, i.e. by following the quantitative test founded by John
Bell to verify the completeness of Quantum Mechanics \cite{3}. For two
correlated spin $\frac{1}{2}$ particles, with spin vectors $\vec{s}_{1}$and $%
\vec{s}_{2}$, the following inequality holds: 
\begin{equation}
-2\leq S=P(\hat{u}_{1};\hat{u}_{2})-P(\hat{u}_{1};\hat{u}_{2}^{\prime })+P(%
\hat{u}_{1}^{\prime };\hat{u}_{2})+P(\hat{u}_{1}^{\prime };\hat{u}%
_{2}^{\prime })\leq 2
\end{equation}
where $\hat{u}_{1}$ and $\hat{u}_{2}$ are unitary norm vectors related to
the angular coordinates $(\Theta _{1},\Phi _{1})$ and $(\Theta _{2},\Phi
_{2})$ in the Bloch sphere and $P(\hat{u}_{1};\hat{u}_{2})$ is the
correlation function 
\begin{equation}
P(\hat{u}_{1};\hat{u}_{2})=\left\langle (\hat{u}_{1}\cdot \vec{s}_{1})(\hat{u%
}_{2}\cdot \vec{s}_{2})\right\rangle \text{.}
\end{equation}
Note that the angle $\Theta $\ on the Bloch sphere corresponds to the
polarization angle $\theta =\Theta /2$. In a density matrix formalism this
last one can be written as: 
\begin{equation}
P(\hat{u}_{1};\hat{u}_{2})=%
\mathop{\rm Tr}%
(\rho O_{12})  \label{1}
\end{equation}
where $O_{12}=O_{1}\otimes O_{2}$,$O_{k}=\left( 
\begin{array}{cc}
\cos \Theta _{k} & e^{-i\,\Phi _{k}}\sin \Theta _{k} \\ 
e^{i\,\Phi _{k}}\sin \Theta _{k} & -\cos \Theta _{k}
\end{array}
\right) $, $k=1,2.$ By choosing $\Theta _{1}=0$, $\Theta _{2}=\pi /4$, $%
\Theta _{1}^{\prime }=\pi /2$, $\Theta _{2}^{\prime }=3\pi /4$, and $\Phi
_{1}=\Phi _{2}=\Phi _{1}^{\prime }=\Phi _{2}^{\prime }=0\ $this leads to the
well known result: $S=2\sqrt{2}>2$ for a spin singlet $\left| \Psi
_{-}\right\rangle .$In the case of MEMS the very general form of density
matrix $\rho $ given by Eq. (8) can be adopted. There $A$, $B$, $C=-%
{\frac12}%
p$, $D$ are real numbers and $D$ depends on the normalization condition, $%
\mathop{\rm Tr}%
\rho =1$: $D=1-A-2B.$ The correlation function becomes 
\begin{equation}
P(\hat{u}_{1};\hat{u}_{2})=(1-4B)\cos \Theta _{1}\cos \Theta _{2}-p\cos
(\Phi _{1}-\Phi _{2})\sin \Theta _{1}\sin \Theta _{2}\text{,}  \label{2}
\end{equation}
where the first term depends on the structure of the state by the diagonal
term $B$ while the second term is function of the only singlet weight $p$ in
the mixture. This leads to the following general expression of the Bell
parameter: 
\begin{eqnarray}
S &=&(1-4B)[\cos \Theta _{1}\cos \Theta _{2}-\cos \Theta _{1}\cos \Theta
_{2}^{\prime }+\cos \Theta _{1}^{\prime }\cos \Theta _{2}+\cos \Theta
_{1}^{\prime }\cos \Theta _{2}^{\prime }]-  \nonumber \\
&&p[\cos (\Phi _{1}-\Phi _{2})\sin \Theta _{1}\sin \Theta _{2}-\cos (\Phi
_{1}-\Phi _{2}^{\prime })\sin \Theta _{1}\sin \Theta _{2.}^{\prime }+ \\
&&\cos (\Phi _{1}^{\prime }-\Phi _{2})\sin \Theta _{1}^{\prime }\sin \Theta
_{2.}+\cos (\Phi _{1}^{\prime }-\Phi _{2}^{\prime })\sin \Phi _{1}^{\prime
}\sin \Phi _{2}^{\prime }].  \nonumber
\end{eqnarray}
The inequality violation conditions are found by looking for an extremal
point of $S$. It can be verified that $%
{\displaystyle{\partial S \over \partial \Theta _{k}}}%
=0$ for: $\cos \Theta _{1}=\cos \Theta _{2}=\cos \Theta _{1}^{\prime }=\cos
\Theta _{2}^{\prime }=0.$ Set $\Theta _{1}=\Theta _{2}^{\prime }=-\pi /2$, $%
\Theta _{1}^{\prime }=\Theta _{2}=\pi /2$ as a convenient choice and, by
substitution, the following expression:

\[
S=p\left[ \cos (\Phi _{1}-\Phi _{2})-\cos (\Phi _{1}^{\prime }-\Phi
_{2})+\cos (\Phi _{1}-\Phi _{2}^{\prime })+\cos (\Phi _{1}^{\prime }-\Phi
_{2}^{\prime })\right] 
\]
attains its minimum value $S=-2\sqrt{2}p$ when $\Phi _{1}=-\Phi _{1}^{\prime
}=\pi /4$, $\Phi _{2}=\pi /2$, $\Phi _{2}^{\prime }=0$. Hence we find that
the violation of the Bell inequalities is observed only for states with $p>1/%
\sqrt{2}$, for any values of the diagonal terms $A,B,D$.

This result holds for Werner states, where $A=D=%
{\frac14}%
(1-p),$ $B=%
{\frac14}%
(1+p)$, as said in Section III.

By a more elegant procedure, we could note that for any traceless operator $%
O_{12}$ the expectation value in a state with weight $p$ is given by 
\begin{equation}
P(\hat{u}_{1};\hat{u}_{2})=%
\mathop{\rm Tr}%
(\rho _{W}O_{12})=pP_{s}(\hat{u}_{1};\hat{u}_{2})\text{.}
\end{equation}
The expression of the Bell inequality then becomes 
\begin{equation}
-\frac{2}{p}\leq P_{s}(\hat{u}_{1};\hat{u}_{2})-P_{s}(\hat{u}_{1};\hat{u}%
_{2}^{\prime })+P_{s}(\hat{u}_{1}^{\prime };\hat{u}_{2})+P_{s}(\hat{u}%
_{1}^{\prime };\hat{u}_{2}^{\prime })\leq \frac{2}{p}\text{,}  \label{4}
\end{equation}
which is violated for $p>1/\sqrt{2}$, i.e. $S_{L}<1/2$. The corresponding
experimental test requires the same angular setting of the $\pi -$analyzer
as any standard Bell test performed for a pure singlet state and reported
in\ Section IIB.

Note that a in the range $p\in \lbrack 
{\displaystyle{1 \over \sqrt{2}}}%
,%
{\displaystyle{1 \over 3}}%
]$, or equivalently in $S_{L}\in \lbrack 
{\displaystyle{1 \over 2}}%
,%
{\displaystyle{8 \over 9}}%
]$, the Werner states, although not separable, do not violate the CHSH
inequality \cite{barrett}.

In summary, for any Werner state, $\rho _{W}\ \ $three regions can be
distinguished:

$%
{\displaystyle{1 \over \sqrt{2}}}%
<p\leq 1$ $\ \ \ \ \ (0\leq S_{L}<%
{\displaystyle{1 \over 2}}%
)$ $\ \ \ \ \ \rho _{W}$ is not separable and violates local realism;

$%
{\displaystyle{1 \over 3}}%
<p\leq 
{\displaystyle{1 \over \sqrt{2}}}%
$ $\ \ \ (%
{\displaystyle{1 \over 2}}%
\leq S_{L}<%
{\displaystyle{8 \over 9}}%
)$ \ \ \ \ $\rho _{W}\ $is not separable and does not violate the CHSH
inequality;

$0\leq p\leq 
{\displaystyle{1 \over 3}}%
$ \ \ $\ \ \ \ \ \ (%
{\displaystyle{8 \over 9}}%
\leq S_{L}\leq 1)$ \ \ \ \ \ \ $\rho _{W}$ is separable and local .

The following corresponding values of the $|S|$ parameter have been obtained
experimentally : $|S|=2.0790\pm 0.0042$ (Fig. 7a), $|S|=1.049\pm 0.011$
(Fig. 7b), $|S|=0.4895\pm 0.0047$ (Fig. 7c).

The test of Bell-CHSH inequalities has been performed for several generated
Werner states. The relative behaviour of $|S|$ as a function of the weight $%
p $ is shown in Fig. 12. The experimental data, placed under the expected
straight line show the effect of a non perfect correlation within the
singlet region ${\bf A}$ in Fig.5.

We may extend the above consideration to any MEMS state, $\rho _{MEMS\;}$,
where $A=1-2g(p)$, $B=g(p)$, $D=0$. \ The correlation function becomes 
\begin{equation}
P(\hat{u}_{1};\hat{u}_{2})=(1-4g(p))\cos \Theta _{1}\cos \Theta _{2}-p\cos
(\Phi _{1}-\Phi _{2})%
\mathop{\rm sen}%
\Theta _{1}%
\mathop{\rm sen}%
\Theta _{2}\text{.}
\end{equation}
In this case the violation, $p=%
{\displaystyle{1 \over \sqrt{2}}}%
$, occurs for the point with coordinates $(S_{L},T)=(0.552,\,0.5).$ An
experimental test of Bell's inequality, performed with these states,
involves observables with a complex phase term $\Phi $.

Two distinct regions can be distinguished in the case of MEMS:

$%
{\displaystyle{1 \over \sqrt{2}}}%
<p\leq 1$ $\ \ \ (0\leq S_{L}<0.552)$ \ \ \ \ \ \ $\rho _{MEMS\;}$is not
separable and violate local realism;

$%
{\displaystyle{1 \over \sqrt{2}}}%
<p\leq 0$ $\ \ (0.552\leq S_{L}<%
{\displaystyle{8 \over 9}}%
)$ \ \ \ \ \ \ $\rho _{MEMS\;}$is not separable and do not violate CHSH
inequality.

\section{Conclusion and acknowledgments}

The novel structural characteristics of the SPDC source of $\pi -$entangled
photon pairs and the new way the quantum superposition principle underlying
its performance is exploited, are expected to suggest in the future an
increasing number of sophisticated applications in the domain of quantum
information, quantum communication and in the broader field dealing with of
the fundamental tests of quantum mechanics. We can think, for instance to
the entanglement multiplexing in quantum cryptographic networks, new schemes
of joint entanglement over $\overrightarrow{\pi }$ and ${\bf k}$-vector
degrees of freedom etc. As already mentioned, adding a spherical cavity to
the new device will allow the realization of a new kind of Optical
Parametric Oscillator (OPO)\ synchronously pumped by a high peak-power,
mode-locked femtosecond UV\ source. This device would generate a multiphoton
non-thermal entangled state over a spatially extended {\it E-ring}
distribution of ${\bf k}$-vectors. As said, this idea is presently being
investigated in our Quantum Optics Laboratory in Roma.

We already emphasized that, as far as applications to quantum-mechanical
foundations are concerned, the present device appears to be an ideal source
of bi-partite {\it mixed-states, }as{\it \ }reported at length in the
present paper{\it , }and of \ bi-partite {\it non-maximally entangled }states%
{\it .}\ In this last respect, a comprehensive study on the verification of
the {\it Hardy's ladder proof, }a{\it \ }relevant quantum{\it \ }nonlocality
test, has been succesfully completed recently \cite{11} and will be reported
extensively elsewhere. There the performance of the source was so good as to
allow the attainment of \ the ladder's rung $K=20$, while previous
experiments were limited to $K\leq 3$ \cite{14,36}.

In summary, we do believe that the present work may represent a real
breakthrough in an important branch of modern science and certainly will
lead soon to relevant applications of advanced technology.

Thanks are due to W.Von Klitzing and G. Giorgi for early involvement in the
experiment and for useful discussions. This work was supported by the FET
European Network on Quantum Information and Communication (Contract
IST-2000-29681: ATESIT) and by PRA-INFM\ 2002 (CLON).

\vskip 8mm

\parindent=0pt

\parskip=3mm

\section{Appendix A}

{\bf Phase Control of the Entangled state}. Because of its peculiar
configuration of single arm interferometer the high brilliance source
overcomes many of the instability problems due to the typical phase
fluctuations of a standard two arm interferometer. The spherical mirror $M$
determines a large value of the displacement, $\left| \Delta d\right| \simeq
60\mu m$ in our case, which allows the phase transition $\phi =0\rightarrow $%
\ $\phi =\pi $ from Bell state $|\Phi _{+}\rangle $ to the other $|\Phi
_{-}\rangle $. This can be demonstrated by referring\ \ to Fig. 13. Starting
from the condition $d=R$ and neglecting for simplicity the BBO and $\lambda
/4$ wp thickness, a displacement $\Delta d=OO\prime =AA\prime \neq 0$
determines different optical paths of the UV beam, $2(OA\prime )$, and of
the photon pair generated toward the left in Fig.13 and reflected by mirror $%
M$, $2(OB\prime +B\prime C)$. The factor $2$ which compares in the second
optical path accounts for the two photon emission over two symmetric
directions with respect ${\bf k}_{p}$. Phase difference has the following
expression: 
\[
\phi =4\pi (OA\prime )\lambda _{p}^{-1}-4\pi (OB\prime +B\prime C)\lambda
^{-1}=4\pi \lambda ^{-1}\left[ 2(OA\prime )-(OB\prime +B\prime C)\right] 
\text{.} 
\]
$\phi $ is function of the distances $OA\prime $, $OB\prime $ and $B\prime C$%
. Since $OA=OB=O\prime A\prime =O\prime B\prime =R$, we have: $OA\prime
=R+\Delta d$. By approximating $\alpha \prime \prime \cong \alpha \prime
\cong \alpha $ and applying the Carnot theorem to the triangle $OO\prime
B\prime $, we find the following expression: $OB\prime =\sqrt{\left( \Delta
d\right) ^{2}+R^{2}+2R\Delta d\cos (\alpha )}$ Finally the following
equality holds: $B\prime C=O\prime \prime C+O\prime \prime B\prime $, and it
can be easily found: $(O\prime \prime B\prime )=-(OO\prime \prime )\cos
(\alpha )+(OB\prime )$, while $O\prime \prime C$ is obtained by simple
calculations.

By the above results we easily find that the transition $|\Phi _{+}\rangle
\rightarrow $ $|\Phi _{-}\rangle $ is obtained by a displacement $\left|
\Delta d\right| =60\mu m$, which is in good agreement with the experimental
results.

Note that the condition $\Delta d\neq 0$ implies a lateral displacement $OC$
of the reflected SPDC beams. Because of the intrinsic cylindrical symmetry, $%
OC$ may be viewed as the radius of an annular region which grows with $%
\Delta d$ on the BBO plane. This geometrical effect makes the two emission
cones distinguishable. It introduces a {\it spatial }decoherence which
becomes relevant as far as $OC$\ becomes comparable with the diameter of the 
{\it active} pumped region of the crystal ($\simeq 150\mu m$). In our
experimental conditions, $OC\simeq 10^{-1}\Delta d$, we have observed that
any coherent superposition on the state vanishes for $\left| \Delta d\right|
\gtrapprox 600\mu m$.

\centerline{\bf Figure Captions}

\vskip 8mm

\parindent=0pt

\parskip=3mm

\begin{description}
\item[Fig. 1 - ]  Layout of the universal{\it , }high-brilliance{\it \ }%
source of \ polarization entangled photon states and of general mixed
states. Inset: Entanglement-ring and pinhole of radius $r$ for spatial
selection.

\item[Fig. 2 - ]  Bell inequalities test. The selected state is $\left| \Phi
_{-}\right\rangle =\frac{1}{\sqrt{2}}\left( \left| H_{1},H_{2}\right\rangle
-\left| V_{1},V_{2}\right\rangle \right) $.

\item[Fig. 3 - ]  Coincidence rate versus the position X of the beam
splitter in the Ou-Mandel experiment. In the inset the corresponding
interferometric apparatus is shown. The selected state is $|\Psi \rangle =%
\frac{1}{\sqrt{2}}\left( |H_{1},V_{2}\rangle +e^{i\phi }|V_{1},H_{2}\rangle
\right) $.

\item[Fig. 4 - ]  Plot of the fringe visibility (cirles, left axis) and
coincidence rate (squares, right axis) as a function of the iris diafragm
radius $r$.

\item[Fig. 5 - ]  Partition of the (half) {\it E-ring} for Werner states
production.

\item[Fig. 6 - ]  Tomographic reconstruction (real parts) of the mixtures $%
\frac{1}{2}\left[ \left| \Phi _{-}\right\rangle \left\langle \Phi
_{-}\right| \text{+}\left| \Psi _{-}\right\rangle \left\langle \Psi
_{-}\right| \right] $ (a) and $\frac{1}{4}{\bf I}_{4}$ (b).

\item[Fig. 7 - ]  Tomographic reconstruction (real parts) of three different
Werner states. Corresponding singlet weights $p$ are also shown.

\item[Fig. 8 - ]  $Tangle$ vs $Linear$ $Entropy$ for experimentally
generated Werner states. Continous line corresponds to the theoretical curve.

\item[Fig. 9 - ]  Partition of the (half) {\it E-ring} for MEMS production:
(a) left cone for $0\leq p\leq 
{\displaystyle{2 \over 3}}%
$, (b) right cone, (c) left cone for $%
{\displaystyle{2 \over 3}}%
\leq p\leq 1$.

\item[Fig. 10 - ]  Tomographic reconstruction (real parts) of two different
MEMS. Corresponding singlet weights $p$ are also shown.

\item[Fig. 11 - ]  $Tangle$ vs $Linear$ $Entropy$ for experimentally
generated MEMS. Continous line represents theoretical behaviour, dotted line
is the expected curve for Werner states.

\item[Fig. 12 - ]  Plot of the Bell parameter $|S|$ as a function of weight $%
p$. The expected straight line $|S|=2\sqrt{2}p$ is also reported.

\item[Fig. 13 - ]  Scheme representing the optical path difference within
the single arm interferometer.
\end{description}

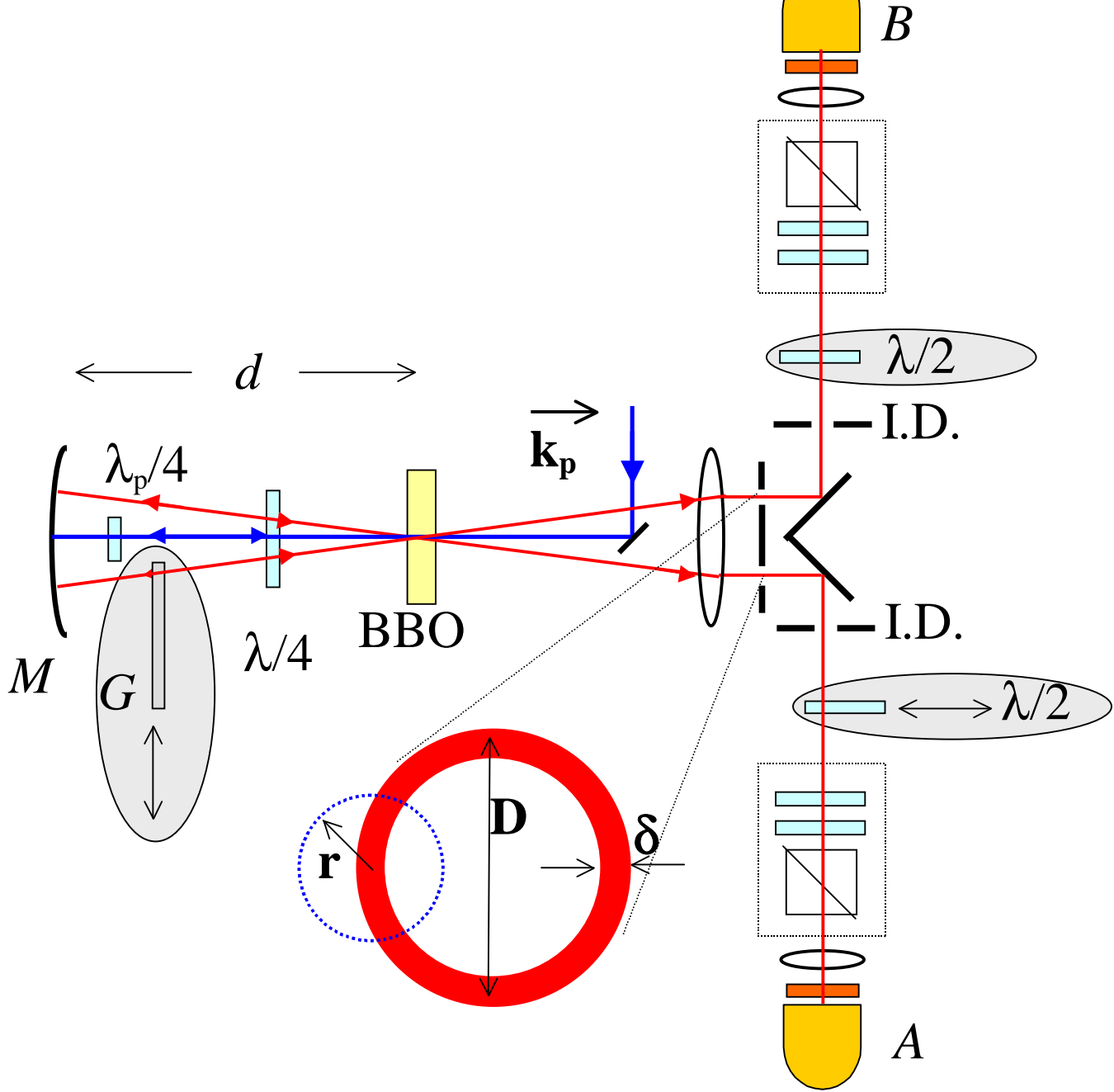
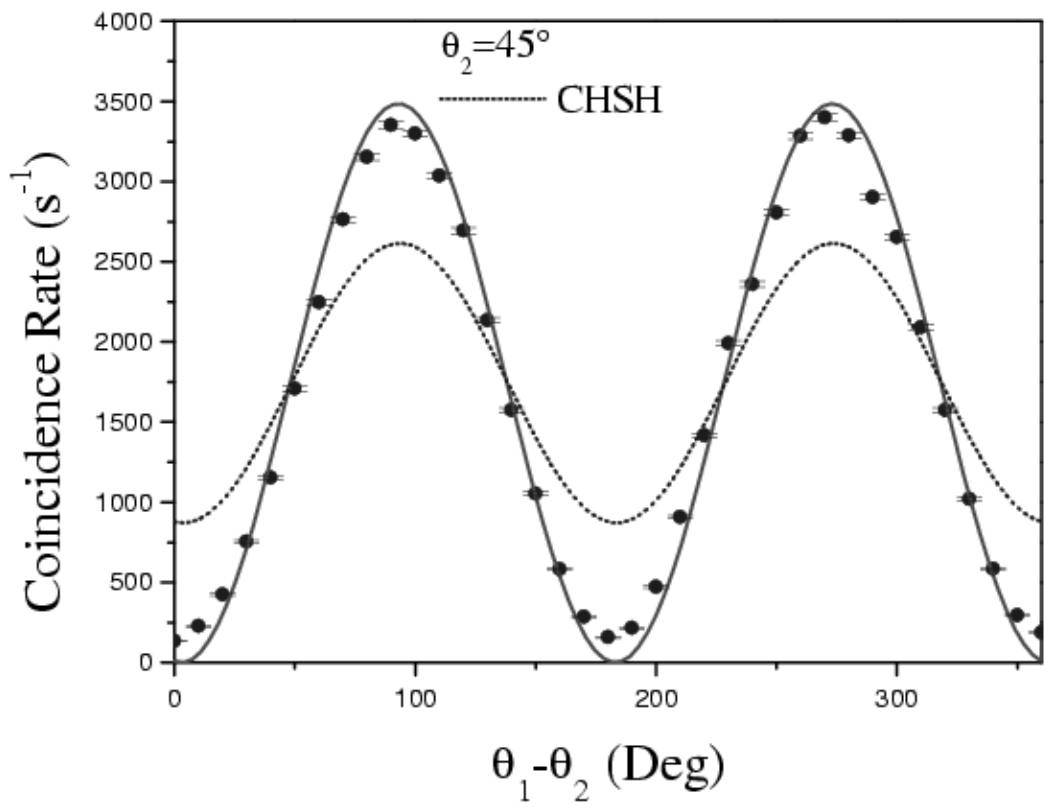
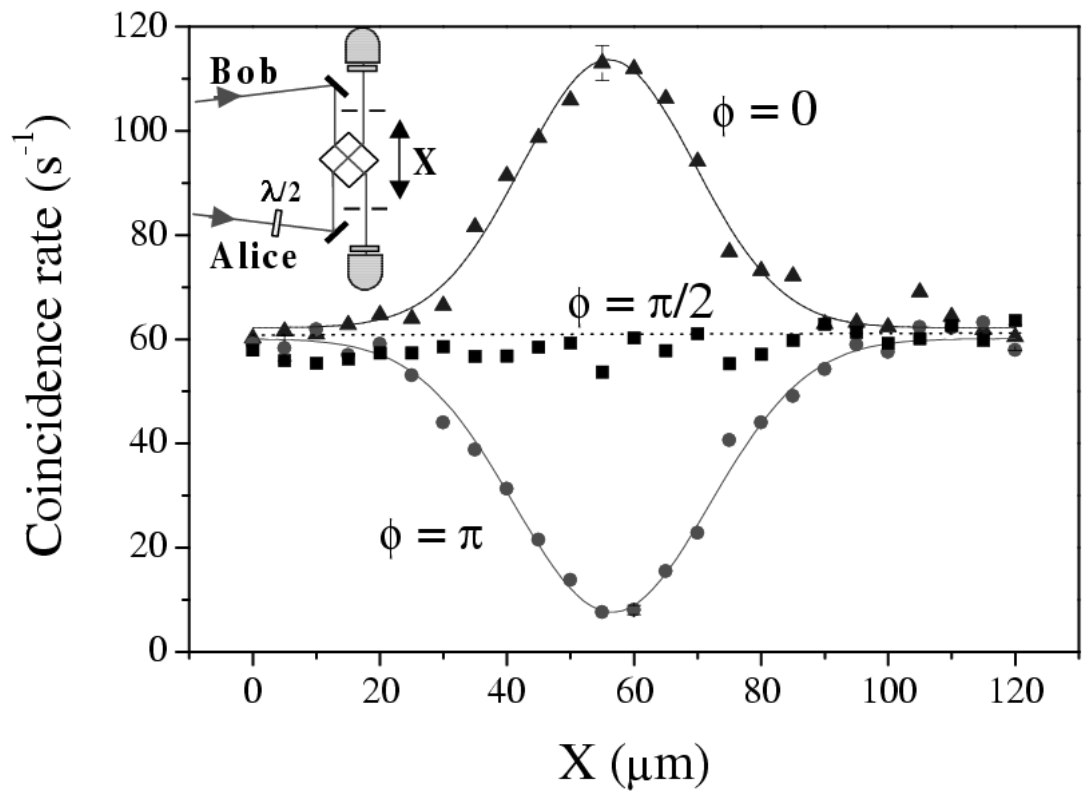
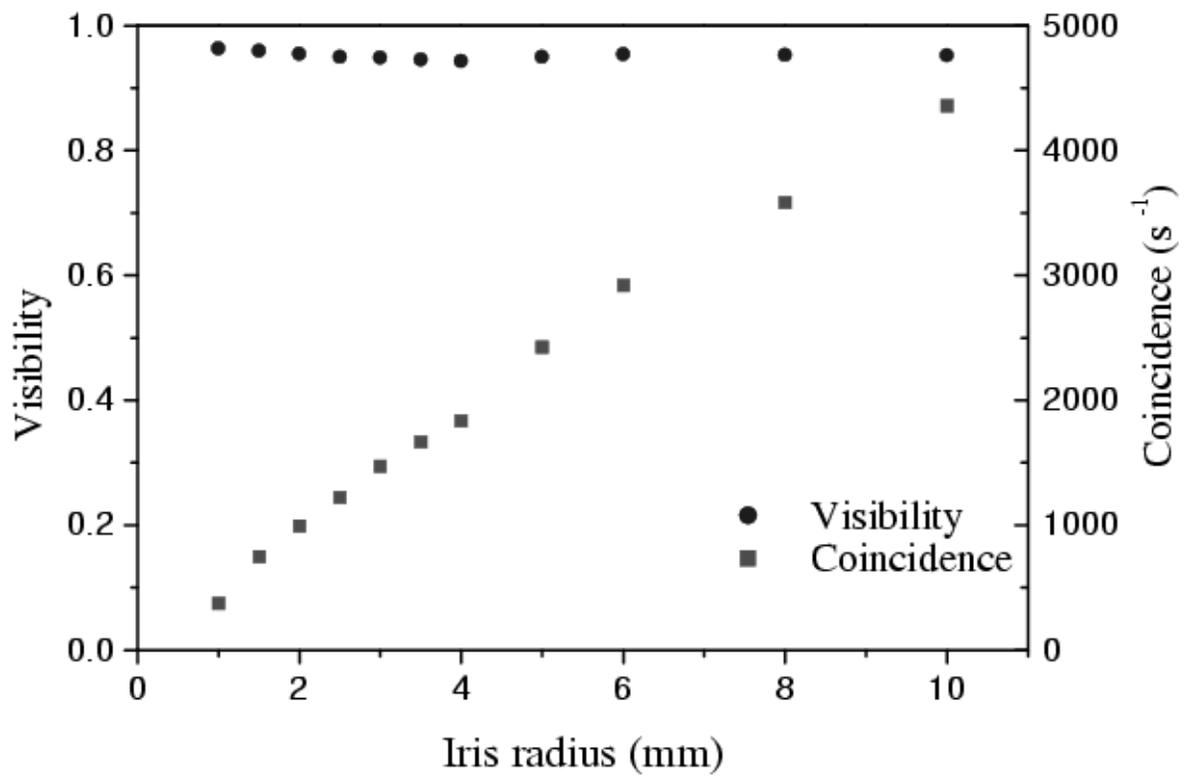
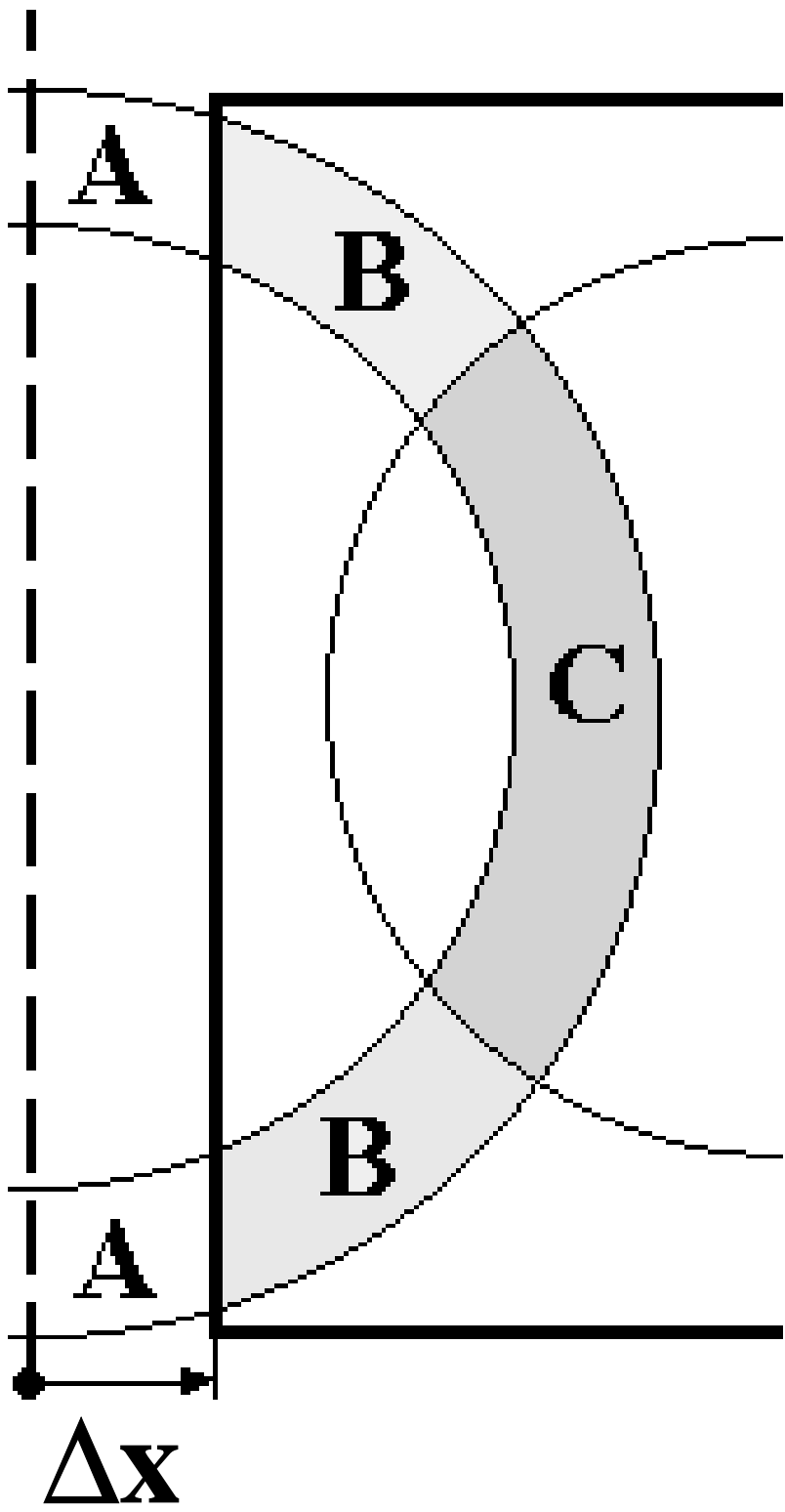
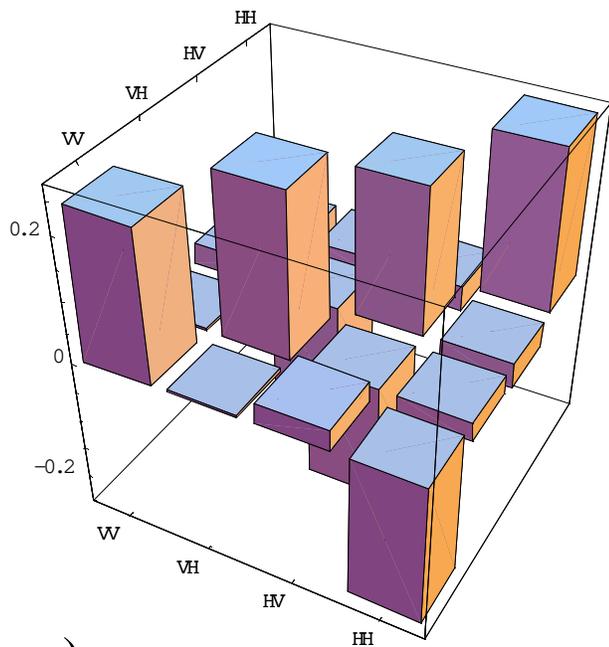
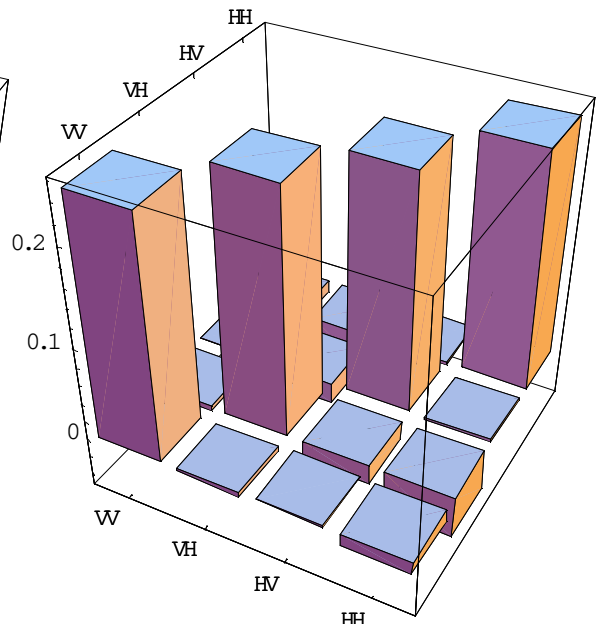
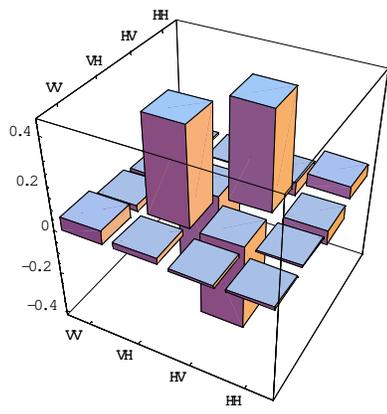
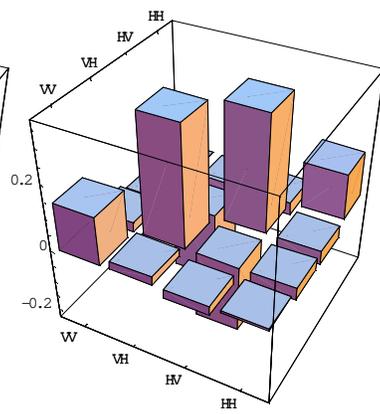
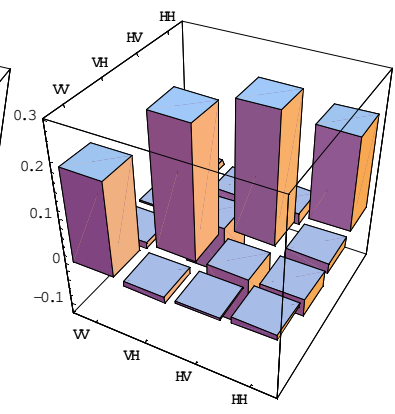
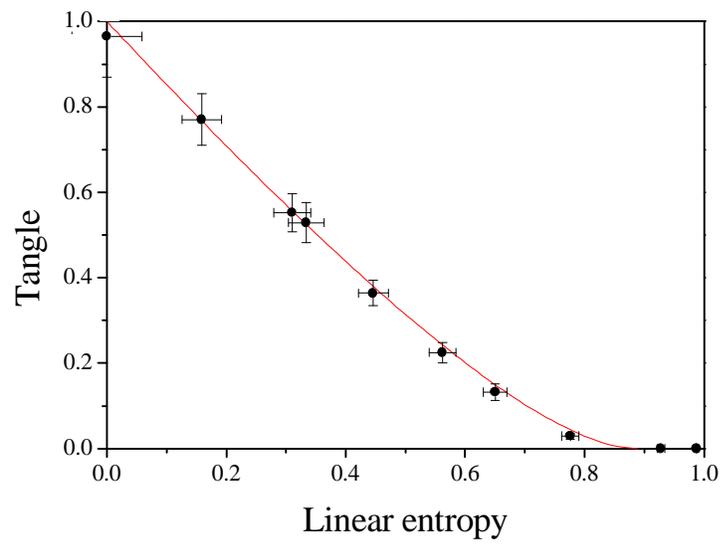
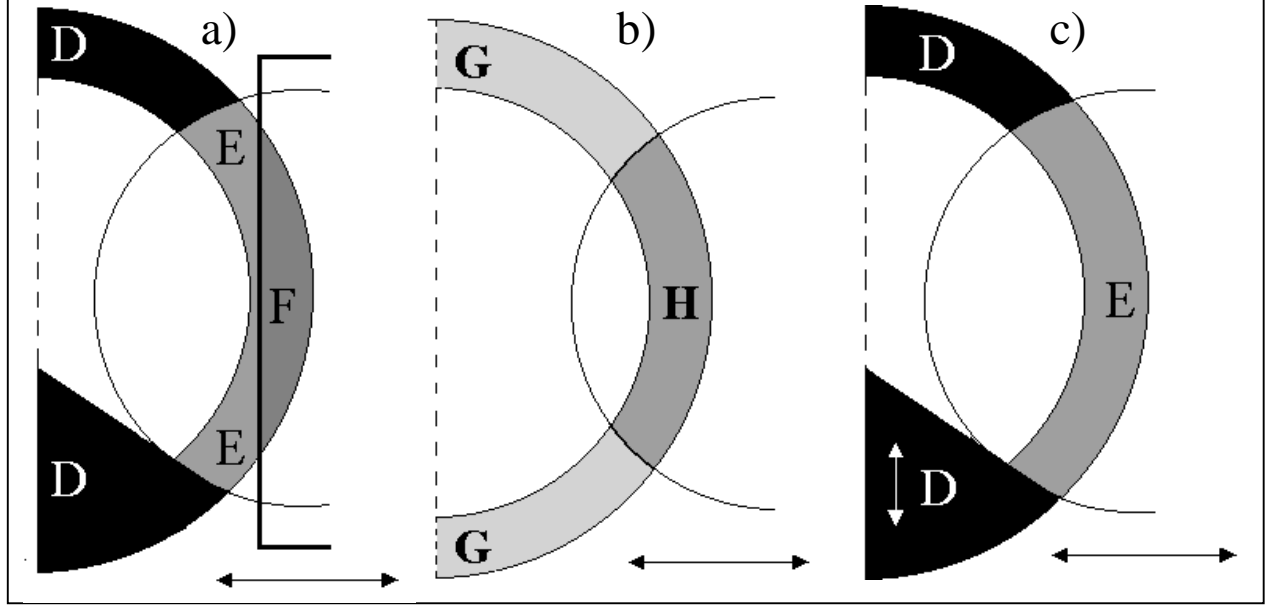
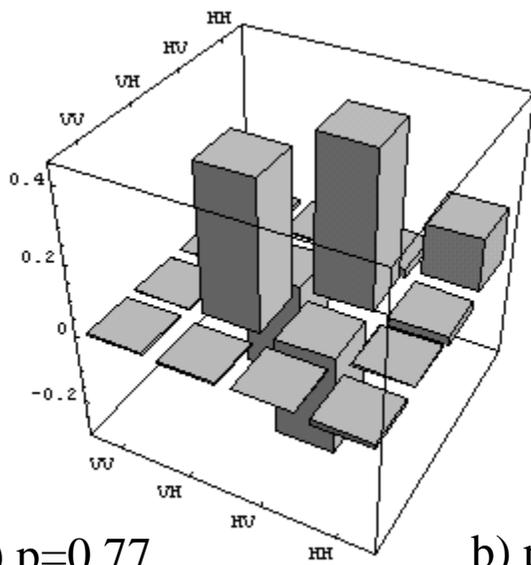
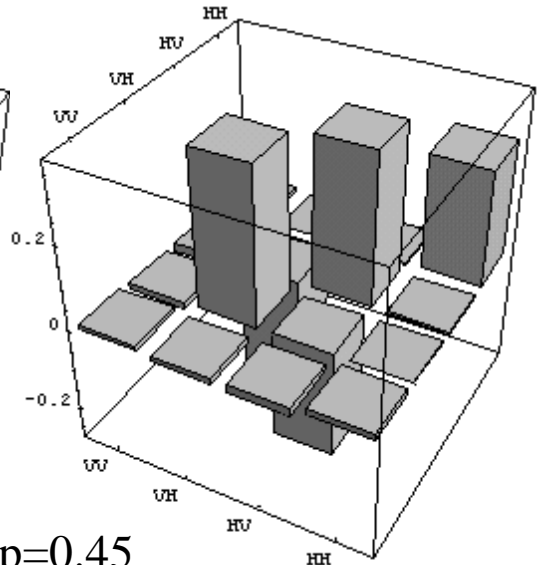
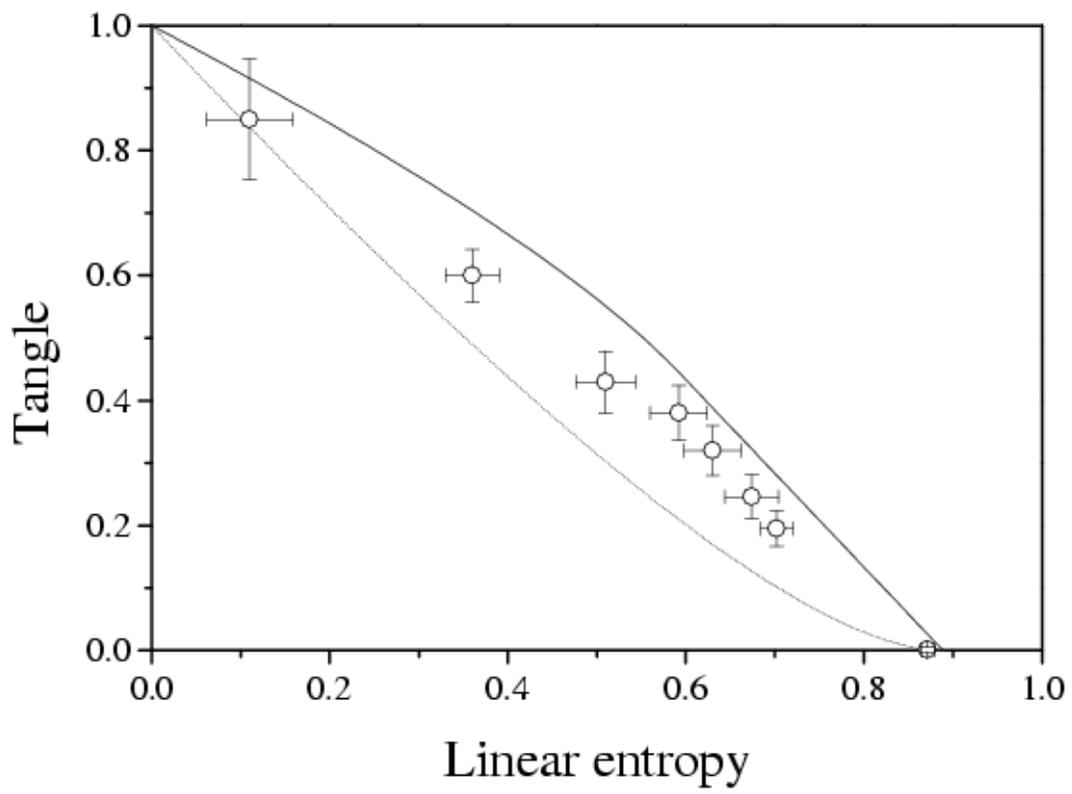
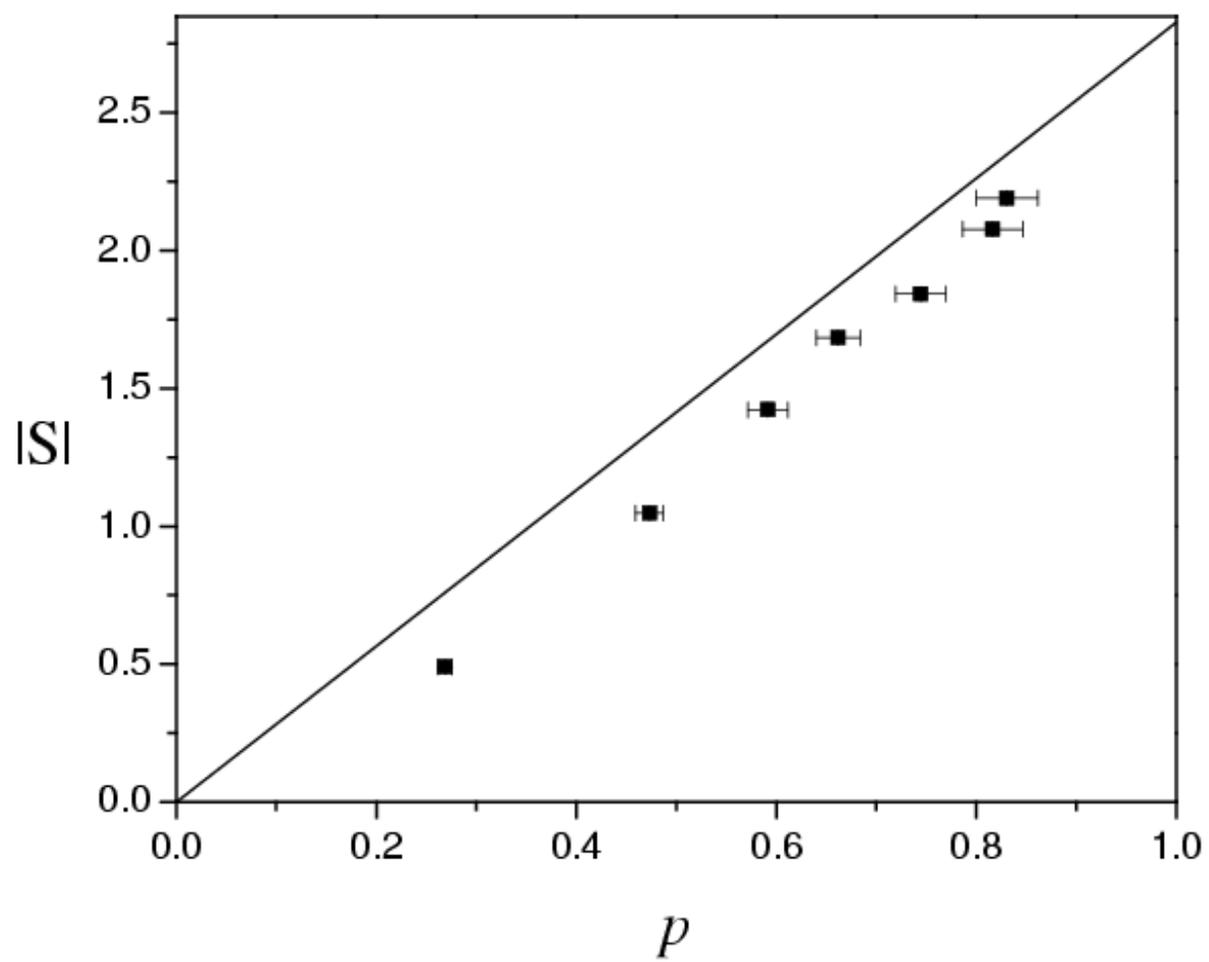
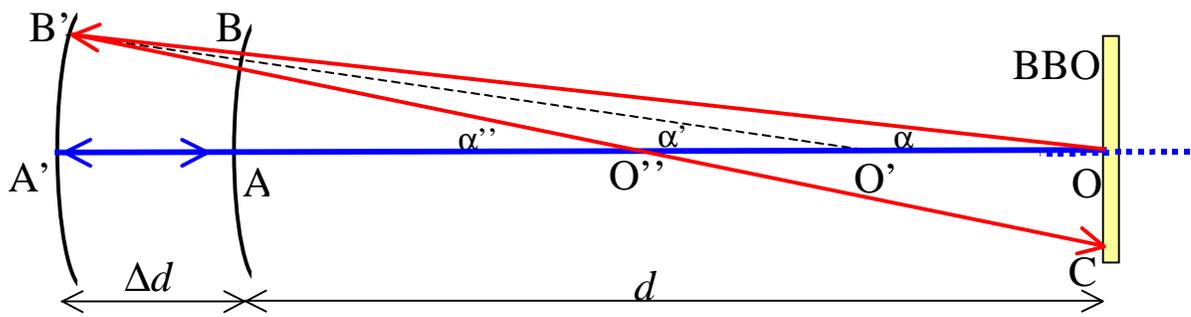
\end{document}